\documentclass[]{interact}

\usepackage{epstopdf}
\usepackage{subfigure}

\usepackage{natbib}
\bibpunct[, ]{(}{)}{;}{a}{}{,}
\renewcommand\bibfont{\fontsize{10}{12}\selectfont}

\theoremstyle{plain}

\theoremstyle{definition}

\theoremstyle{remark}

\usepackage{url} 
\usepackage{bookmark}
\usepackage[ruled]{algorithm2e}
\usepackage{amsmath,bm}

\begin{document}


\title{Impact Evaluation of Falsified Data Attacks on Connected Vehicle Based Traffic Signal Control}

\author{
\name{Shihong~Ed Huang\textsuperscript{a}, Wai Wong\textsuperscript{a}, Yiheng Feng\textsuperscript{b},Qi Alfred Chen\textsuperscript{c}, Z. Morley Mao\textsuperscript{d}, and Henry X. Liu\textsuperscript{a,e}\thanks{CONTACT Yiheng Feng. Email: feng333@purdue.edu}}
\affil{\textsuperscript{a}Department of Civil and Environmental Engineering, University of Michigan, Ann Arbor, MI 48109, USA;
\textsuperscript{b}Lyles School of Civil Engineering, Purdue University, West Lafayette, IN 47907, USA;
\textsuperscript{c}Department of Computer Science, University of California, Irvine, CA 92617, USA; 
\textsuperscript{d}Department of Electrical Engineering and Computer Science, University of Michigan, Ann Arbor, MI 48109, USA;
\textsuperscript{e}University of Michigan Transportation Research Institute, University of Michigan, Ann Arbor, MI 48109, USA}
}

\maketitle

\begin{abstract}
Connected vehicle (CV) technology enables data exchange between vehicles and transportation infrastructure and therefore has great potentials to improve current traffic signal control systems. However, this connectivity might also bring cyber security concerns. As the first step in investigating the cyber security of CV-based traffic signal control (CV-TSC) systems, potential cyber threats need to be identified and corresponding impact needs to be evaluated. In this paper, we aim to evaluate the impact of cyber attacks on CV-TSC systems by considering a realistic attack scenario in which the control logic of a CV-TSC system is unavailable to attackers. Our threat model presumes that an attacker may learn the control logic using a surrogate model. Based on the surrogate model, the attacker may launch falsified data attacks to influence signal control decisions. In the case study, we realistically evaluate the impact of falsified data attacks on an existing CV-TSC system (i.e., I-SIG).
\end{abstract}


Word count: 8132

\section{Introduction}
We are moving into the era of the Internet of Things, in which almost all physical devices will be connected and exchanging data. The same trend is emerging in transportation systems, wherein vehicles and infrastructure are being connected through advanced wireless communication technology, such as Dedicated Short-Range Communications (DSRC) \citep{kenney2011dedicated} or Cellular Vehicle-to-Everything (C-V2X) \citep{wang2017overview}. Connected Vehicles (CVs) are capable of communicating with other CVs through Vehicle-to-Vehicle (V2V) communication and with infrastructure through Vehicle-to-Infrastructure (V2I) communication. The adoption of CV technology has gradually become a national effort, supported by the U.S. Department of Transportation (USDOT), state DOTs, and local DOTs. For instance, three CV pilot programs at New York City, Tampa, and Wyoming have been launched by USDOT in 2016 to deploy and test CV-based applications \citep{CVpilot}.

CVs have great potential to improve a variety of mobility applications, including traffic signal control (TSC), a critical component in urban traffic operation. CV-based TSC (CV-TSC) systems usually utilize CV trajectories as the data source and can allocate green time more efficiently to accommodate real-time traffic conditions. In terms of both cost and performance, CV-TSC systems have significant improvement over conventional TSC systems, which mostly rely on fixed-location sensors (e.g., loop detectors, cameras) for data collection and decision making. Because of these benefits, it is envisioned that CV-TSC systems may gradually replace the conventional TSC systems in the future \citep{MMITSS}. Over the past decade, various CV-TSC models have been proposed to optimize signal timing plans based on real-time CV data \citep{goodall2013traffic,feng2015real,he2012pamscod,priemer2009decentralized,he2014multi,lee2013cumulative,guler2014using,pandit2013adaptive,li2018connected,beak2017adaptive,feng2018real,zheng2018traffic,yu2018integrated,feng2018spatiotemporal,yu2019corridor,yang2019eco}.  

Despite great improvement in system performance, the new trait — connectivity between vehicles and infrastructure — might open a new door to cyber attacks. The benefits of CV-TSC systems can be achieved only if the systems are secure in cyberspace. Therefore, cyber security is a crucial component when developing CV-TSC systems. However, most of the existing CV-TSC models are designed without considering cyber security issues and thus may be vulnerable to cyber attacks. Before developing defense strategies, it is important to identify potential cyber threats and investigate the impact of cyber attacks in terms of system performance. Existing studies on this topic usually consider a `white-box' attack scenario, which assumes that attackers have full access to the traffic control system and/or control logic so that they can manipulate the traffic signal phasing and timing freely \citep{ghena2014green,laszka2016vulnerability,ghafouri2016vulnerability,feng2018vulnerability,chen2018exposing,yen2018falsified,perrine2019implications}. For example, \citet{perrine2019implications} assumes that the traffic signals can be selectively disabled to flashing-red status (equivalent to a four-way stop-sign intersection) without explaining how to make this happen. \citet{chen2018exposing} assumes that the source code of the signal control model is known to the attacker so a comprehensive analysis can be performed. However, this is not a common setting. The major drawback of the white-box attack scenario is that it cannot realistically evaluate the impact of cyber attacks. Although it can provide the upper bound of the system impact under cyber attacks, `manipulating signal phasing and timing freely' is a very strong and unrealistic assumption. For most of the commercial traffic control systems, various levels of protection from hardware (e.g., Malfunction Monitoring Unit) to software (e.g., controller firmware) are designed to prevent such manipulations. In fact, the confidentiality and complexity of the control logic themselves imply certain levels of system security. Thus, it is necessary to realistically evaluate the impact of cyber attacks under real-world scenarios where the controller hardware and control logic are inaccessible. 

In this paper, we aim to evaluate the impact of cyber attacks on CV-TSC systems by considering a more realistic attack scenario. We assume that the attackers do not know the details of the signal control system and do not have physical access to the system (e.g., signal control cabinet). Hence, we name our study as `black-box' attack scenario. Attackers are assumed to learn the signal control logic using a surrogate model. With the learned model, the signal timing plan can be predicted based on critical traffic features that are obtained from real-time traffic conditions. The critical traffic features and subsequently the signal timing plan can be manipulated by attackers who may inject falsified CV data into the system. The impact of cyber attacks is measured by comparing the system performance with and without attacks. In the case study, we realistically evaluate the impact of cyber attacks on I-SIG, an existing CV-TSC system. Results show that falsified data attacks can create excessive delay at the intersection and degrade the performance of the CV-TSC system. We also briefly discuss two defense strategies.

The rest of this paper is organized as follows. The background of CV-TSC systems and related studies regarding cyber attacks are introduced in Section \ref{background}. The threat model, which defines the operating environment and attackers' capabilities, is presented in Section \ref{threatmodel}. Section \ref{learning} assumes that attackers may learn the signal control logic using a surrogate model. Section \ref{casestudy} presents a comprehensive case study, which realistically evaluates the impact of cyber attacks on I-SIG. Finally, conclusions are drawn in Section \ref{conclusion}.

\section{Background}\label{background}
In this section, we will first give a brief description of a general CV-TSC system and then introduce related studies regarding cyber attacks on traffic signal control systems.

\subsection{Connected Vehicle Based Traffic Signal Control}

\begin{figure}
\centering
\includegraphics[scale=0.8]{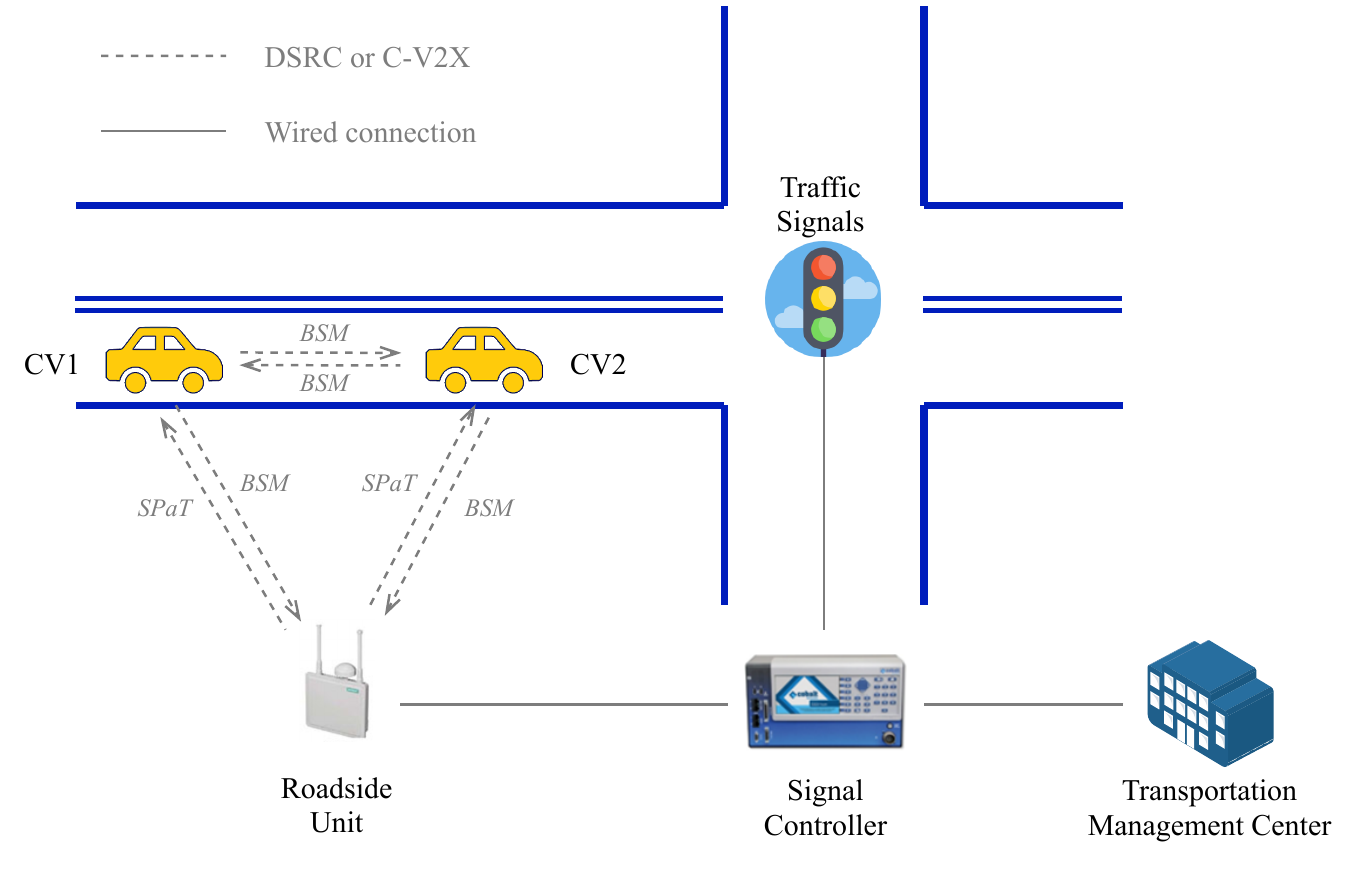}
\caption{Illustration of a connected vehicle based traffic signal control system}
\label{intersection}
\end{figure}

A typical CV-TSC system is illustrated in Figure \ref{intersection}. Each CV is equipped with an On-Board Unit (OBU), which broadcasts Basic Safety Messages (BSMs). A BSM records a CV’s information such as speed, location, heading, and acceleration. Consecutive BSMs represent the vehicle trajectory. On the infrastructure side, an intersection is equipped with a Roadside Unit (RSU), a signal controller, and traffic signals. The RSU receives BSMs, which are used as inputs to optimize traffic signal timing plans. In this paper, we consider the RSU not only can transmit communication messages, but also has computational powers to run signal optimization models and generate optimal signal timing plans. The signal controller executes the optimal signal timing plans and controls the traffic signals to display corresponding colors. The signal controller is connected with a transportation management center (TMC), which can remotely send commands to the signal controller. Meanwhile, the RSU continuously broadcasts Signal Phase and Timing (SPaT) messages, which record current signal status (i.e., green/yellow/red) and remaining time. The CV-TSC system responds to real-time traffic demand and updates the signal timing plans over time based on continuously received BSMs.

\subsection{Related Work}\label{typeofattack}
Based on the general structure of a CV-TSC system in Figure \ref{intersection}, three attack surfaces can be identified at a connected intersection: the transportation infrastructure (RSU, signal controller, and traffic signals), the TMC, and the CVs. Usually, the transportation infrastructure is deployed in an agency's local network with firewall protection. It is difficult to access the infrastructure devices remotely from the public domain. In order to initiate an attack, attackers have to open up the signal control cabinet and establish a wired connection to the devices. Once attackers gain access to the system, they can directly manipulate the signal timing. Alternatively, attackers may trespass into the TMC and send control commands to the signal controller from the TMC. These attacks are referred to as direct attacks, which have been investigated by previous studies \citep{ghena2014green,ernst2017framework,laszka2016vulnerability,reilly2016creating,perrine2019implications}. These studies mainly focus on conventional traffic control systems, for instances, fixed-timing signal control \citep{laszka2016vulnerability} and ramp metering control \citep{reilly2016creating}. Direct attacks, which require conspicuous physical access, can be easily discovered.

The other type of attack is called indirect attacks, in which attackers try to influence signal control decisions by injecting falsified data. Indirect attacks can be launched from the vehicle side. By exploiting software vulnerabilities, it is practically feasible that attackers hack into the communication devices in their own CVs and broadcast falsified BSMs. This is similar to compromising other Electronic Control Units as demonstrated in the literature \citep{koscher2010experimental,checkoway2011comprehensive}. Alternatively, attackers may hack into their vehicles' internal networks. This can be achieved in many ways. For example, attackers may hack into the infotainment system of the vehicle \citep{mazloom2016security}. Once attackers are in the vehicle's internal network, they would be able to take control of a wide range of vehicle functions \citep{koscher2010experimental}, including sending malicious BSMs that contain falsified data elements. Because attackers have arbitrary access to their own vehicles, indirect attacks are much more achievable. Recent studies have shown that falsified input data can indeed influence system control decisions and significantly downgrade the system performance. For example, influencing routing decisions in social navigation systems by generating virtual traffic jams (e.g., Google map, Waze) \citep{jeske2013floating,sinai2014exploiting}, or increasing total travel delay by affecting signal control decisions \citep{ghafouri2016vulnerability,feng2018vulnerability,chen2018exposing,yen2018falsified,oza2020secure}. Compared to direct attacks, indirect attacks are more realistic, and thus are the focus of this study.

\section{Threat Model}\label{threatmodel}

For cyber security research, establishing a realistic threat model is an important prerequisite. A threat model defines an attacker's capabilities, as well as the most relevant and realistic threats to the system. For a CV-TSC system, falsified data attack (i.e., indirect attack) is typically considered as the most plausible type of attack.

\begin{figure}
\centering
\includegraphics[scale=0.8]{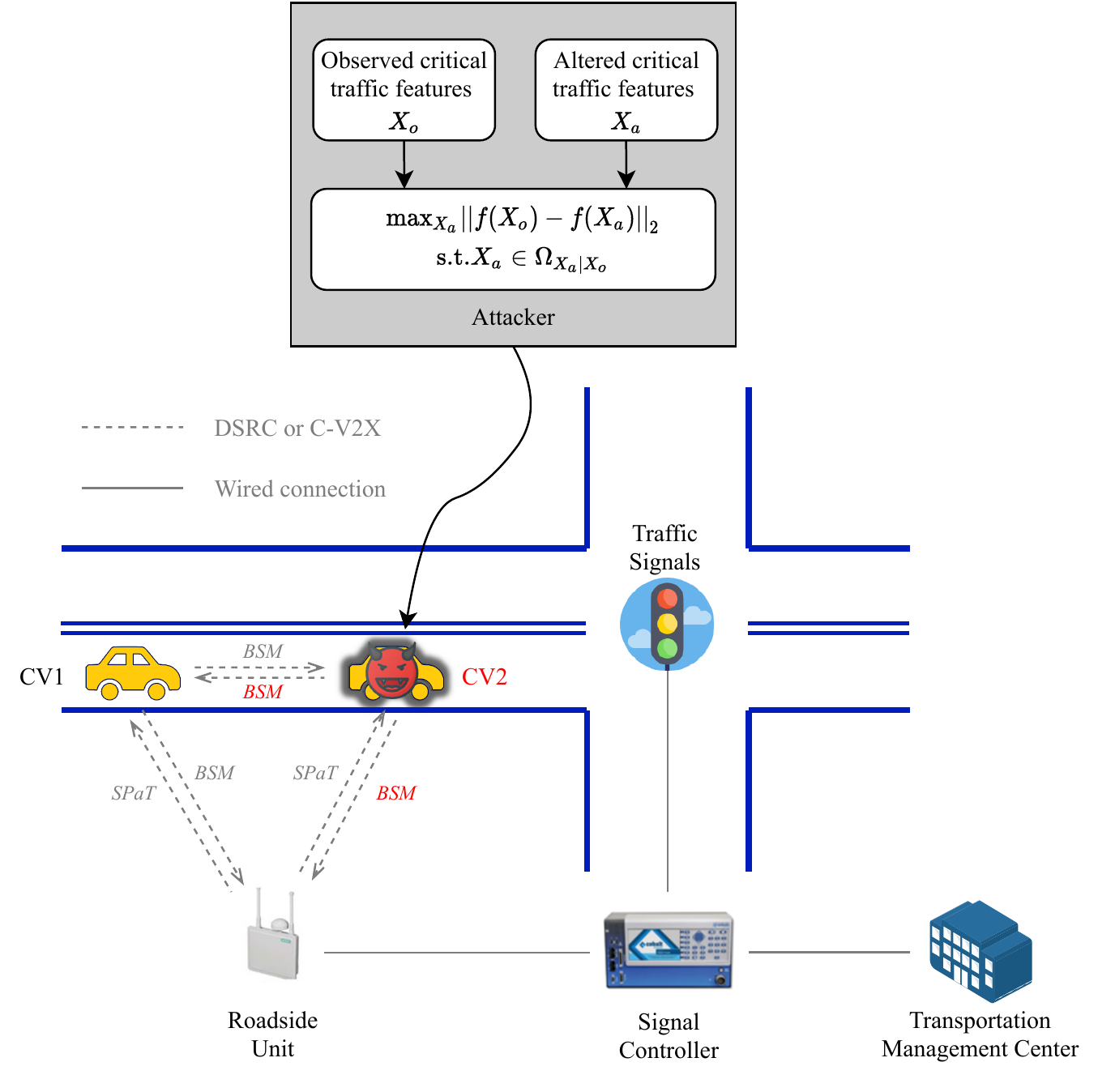}
\caption{Illustration of the threat model}
\label{threat}
\end{figure}

Figure \ref{threat} illustrates the threat model considered in this study. It is similar to Figure \ref{intersection}, except that $\textrm{CV}_2$ is now assumed to be compromised by an attacker. Because all the communications in the vehicular network are in broadcast mode, the attacker receives the same information that other CVs would receive. After analyzing the received BSMs (i.e., trajectories) and SPaT messages (i.e., signal timing), the attacker may `learn' the control logic using a surrogate model $\boldsymbol{f}(\cdot)$. The surrogate model predicts the signal timing plan based on critical traffic features $\boldsymbol{X}$ that are obtained from BSMs (the details on the critical traffic features and surrogate model are introduced in Section \ref{learning}).

Then the attacker may launch cyber attacks by broadcasting falsified BSMs using the compromised communication device (i.e., OBU). Based on the received BSMs, the attacker may evaluate the observed critical traffic features $\boldsymbol{X_o}$ (e.g., queue length) and use the surrogate model to predict the signal timing plan, i.e., $\boldsymbol{f}(\boldsymbol{X_o})$. We refer to $\boldsymbol{f}(\boldsymbol{X_o})$ as the pseudo-optimal timing plan because it is not the exact timing plan generated from the actual control logic, but a plan predicted by the trained surrogate model. By injecting falsified BSMs, the attacker tries to alter the values of the critical features from $\boldsymbol{X_o}$ to $\boldsymbol{X_a}$. Similarly, the attacker may predict the signal timing plan with the altered critical feature, i.e., $\boldsymbol{f}(\boldsymbol{X_a})$. The dissimilarity between the pseudo-optimal timing plan and the timing plan under attack can be computed using the L2 norm. The attacker's goal is to maximize the dissimilarity by generating falsified BSMs that can alter the values of the critical features, as shown in the following problem (\textbf{P1}):
\begin{align}
    \textbf{P1:} \quad & \nonumber \\
    \max_{\boldsymbol{X_a}} \quad &  \|\boldsymbol{f}(\boldsymbol{X_o})-\boldsymbol{f}(\boldsymbol{X_a})\|_2 \label{p1_1}\\
    \textrm{s.t.} \quad & \boldsymbol{X_a} \in \Omega_{\boldsymbol{X_a} | \boldsymbol{X_o}} \label{p1_2}
\end{align}
In \textbf{P1}, the feasible region $\Omega_{\boldsymbol{X_a} | \boldsymbol{X_o}} $ is dependent on $\boldsymbol{X_o}$. For example, after injecting a falsified stopped vehicle (a falsified vehicle has a legitimate trajectory in the form of BSMs but is not physically on road), the new queue length cannot be smaller than the originally observed queue length. 

The falsified BSMs are mixed with the BSMs from regular CVs. The RSU collects all the BSMs and uses them as input data for real-time traffic signal optimization. The generated signal timing plans are influenced by the falsified BSMs, and thus are no longer optimal. Vehicles spend extra time passing the intersection and hence the total travel time is increased. Our threat model aligns with a recent study on black-box attacks against unknown machine learning models \citep{papernot2017practical}. By using a surrogate model, the attacker in that study crafts adversarial images to fool a target model so that the target model would output erroneous predictions. 

It should be noted that falsified data attacks may only influence the timing (e.g., green time and/or sequence) of the traffic signals. The phase configurations remain the same. For example, the attacker is not able to change the minimum and maximum green time of each phase, turn the intersection into flashing-red status, or enable conflicting phases to be green at the same time. This is because the phase configurations are protected by the Malfunction Monitoring Unit and Cabinet Monitor Unit in the signal control cabinet. 

We claim that this threat model is realistic for three reasons. First, it is practically feasible. As discussed in the previous section, the attacker may hack into his/her own CV and OBU to send falsified BSMs. This is similar to spoofing location data on private phones when playing Pok{\'e}mon Go \citep{zhao2017location}. 

Second, this threat model considers falsified data attacks, which are difficult to be perceived. The attacker uses a legitimate communication device to transmit falsified BSMs with valid digital certificates. This means that the attacker does not need to spoof sender's identity, which is protected by the Security and Credential Management System (SCMS) \citep{whyte2013security,brecht2018security}, but only modifies the contents of the messages (e.g., speed and location data). Falsified BSMs can be correctly signed to pass the SCMS identity check when they are received by the RSU and utilized for signal timing optimization. In other words, the SCMS cannot detect such falsified data attacks.

Third, this threat model minimizes the chance of exposure. In reality, the attacker not only wants to launch effective attacks, but also wants to minimize the possibility of being identified. Falsified data attacks do not require the attacker to physically access transportation infrastructure or remotely connect to the agency's internal network. The attacker only needs to compromise a CV to broadcast falsified messages. Moreover, the compromised CV ($\textrm{CV}_2$ in Figure \ref{threat}) does not need to stay on the road to launch attacks, as long as it is within the communication range of the RSU. 

In the following section, we will detail the training of the surrogate model to learn the control logic and the selection of the critical traffic features to launch effective attacks. Given the primary objective of the current study is to evaluate the impact of cyber attacks on CV-TSC systems, we will focus on the investigation of how the altered critical traffic features impact the signal timing plan. The falsified trajectory generation mechanisms adopted by the attacker are beyond the scope of this study.

\section{Learning Control Logic}\label{learning}
\begin{figure}
\centering
\includegraphics[scale=0.6]{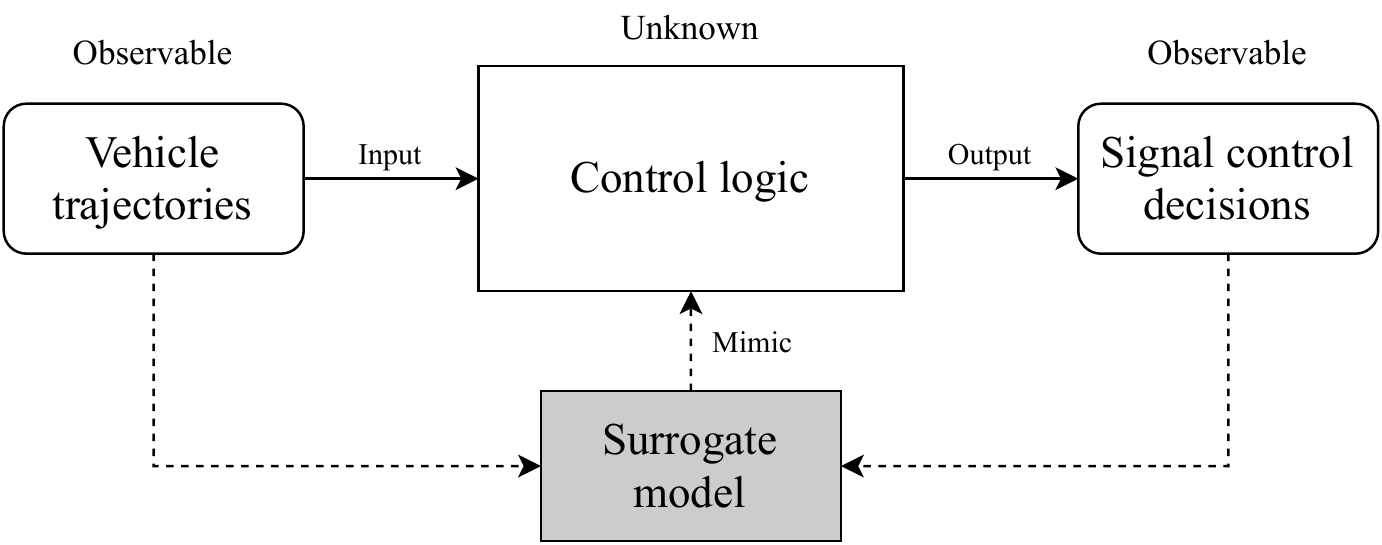}
\caption{The process of learning control logic}
\label{step1}
\end{figure}

In a real-world implementation, the CV-TSC system utilizes vehicle trajectories (obtained from BSMs) as the input and generates optimal signal control decisions. Both vehicle trajectories and signal control decisions are observable to the attacker in the forms of BSMs and SPaT messages respectively. The actual signal control logic, however, is unknown to the attacker. The attacker may adopt a surrogate model to learn the signal control logic. The surrogate model takes the same trajectories as the input and outputs predicted signal timing plan. Historical BSM and SPaT data can be used to train the surrogate model. The attacker uses the surrogate model as the replacement of the real control logic when launching attacks. The whole learning process is illustrated in Figure \ref{step1}. 

\subsection{Traffic Signal Setting}
\begin{figure}
\centering
\includegraphics[scale=0.5]{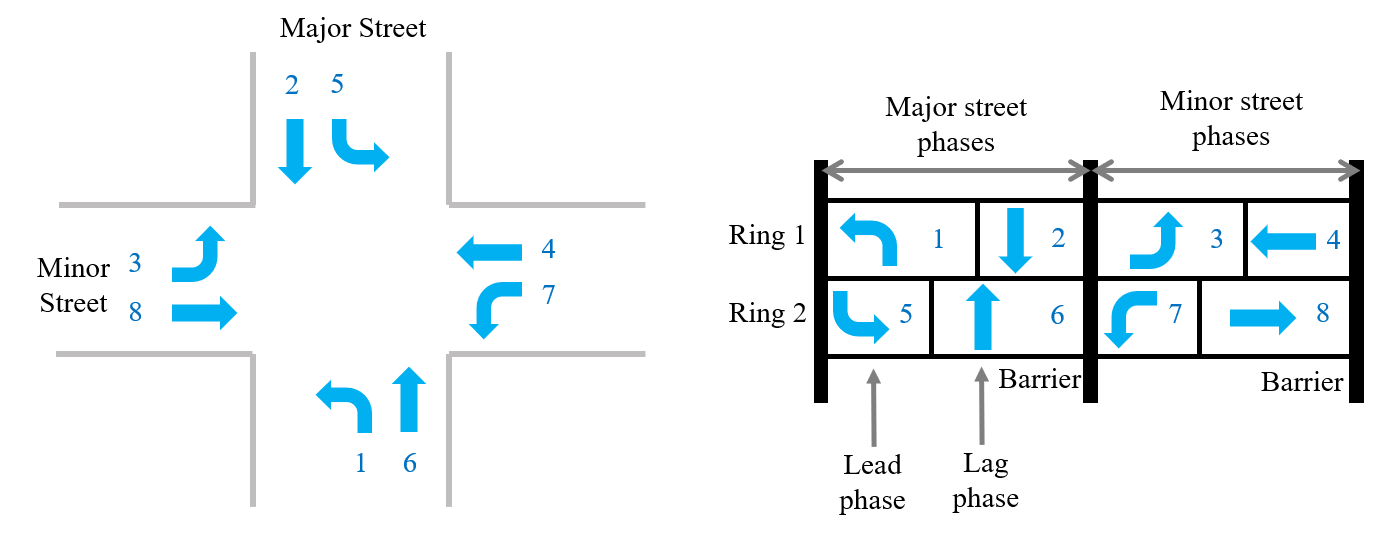}
\caption{Illustration of the ring-barrier structure}
\label{ring-barrier}
\end{figure}

This study assumes that the CV-TSC system uses a ring-barrier phasing. The ring-barrier structure \citep{koonce2008traffic} illustrated in Figure \ref{ring-barrier} is the standard traffic signal setting in North America. Starting from the major street and moving clockwise, the through phases are labeled as phases 2, 4, 6, and 8. Starting from the left-turn phase that is next to phase 6 and moving clockwise, the left-turn phases are labeled as phases 1, 3, 5, and 7. Ring 1 includes phases 1 to 4 and ring 2 includes phases 5 to 8. A barrier separates major street phases (phases 1, 2, 5, and 6) from minor street phases (phases 3, 4, 7, and 8). A barrier may also refer to the four phases of the major street or the four phases of the minor street. The phase that operates first within a barrier is called lead phase and the other one is called lag phase, therefore a barrier includes two lead phases and two lag phases. Usually the signal optimization algorithms change phase sequence and allocate green time of each phase to minimize/maximize predefined performance indexes.

\subsection{Surrogate Model}\label{sec_tree}
A signal timing plan includes two parts, green time of each phase and phase sequence. The prediction of green time can be considered as a regression problem because green time is continuous. In contrast, the prediction of phase sequence is a classification problem since there is a finite number of possibilities for phase sequences. In this study, decision tree regression/classification \citep{breiman2017classification} is adopted to be the surrogate model that could potentially be leveraged by the attacker. Decision tree models are chosen because they are easy to implement and their output always falls within the feasible ranges, i.e., minimum and maximum green time. Most importantly, decision tree models possess inherent `if-then-else' structures and can effectively map nonlinear relationships, making signal control algorithms particularly easy to fit into programmatic structures.

\begin{figure}
\centering
\includegraphics[scale=0.26]{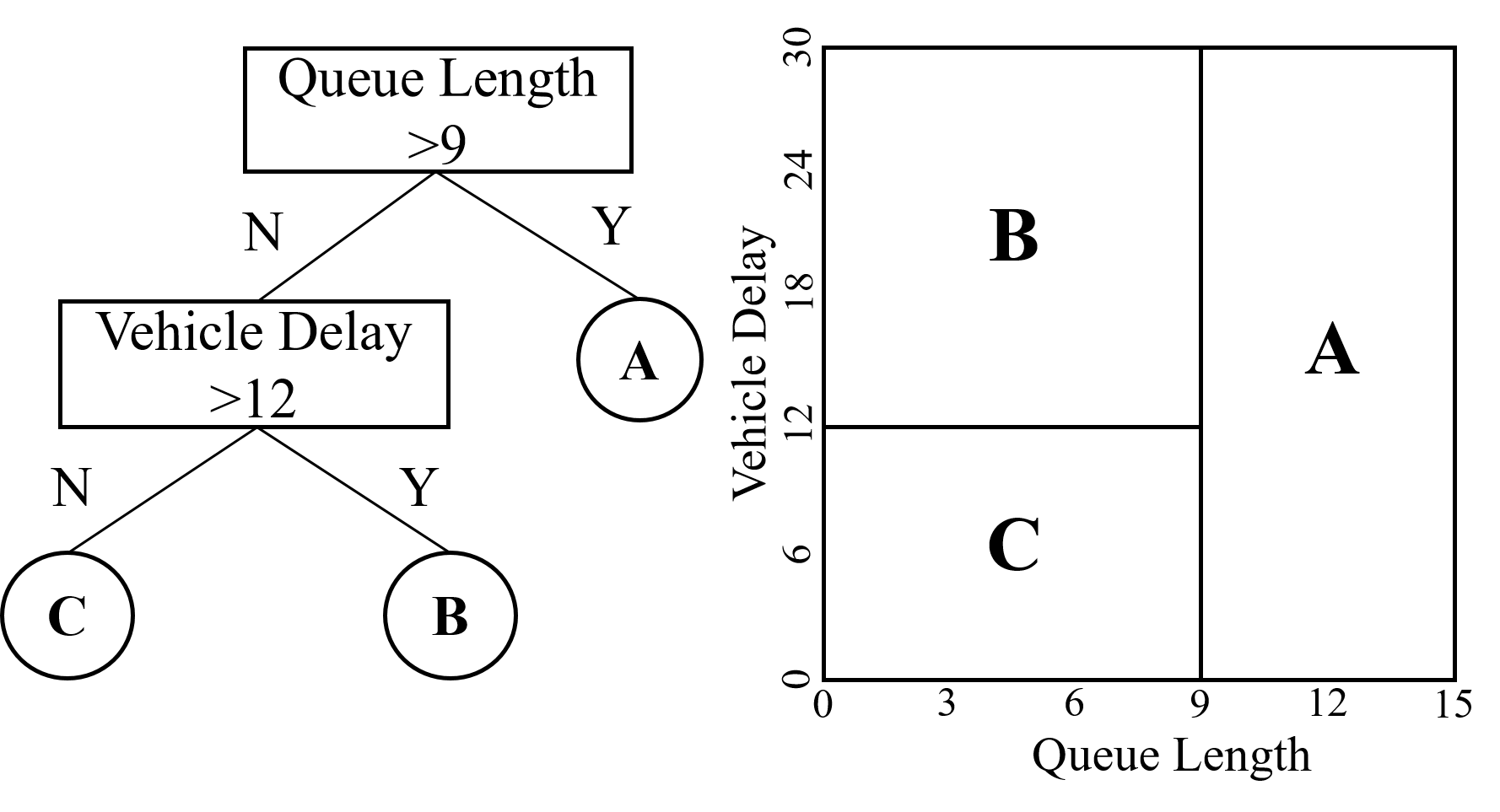}
\caption{Illustration of decision tree model}
\label{tree}
\end{figure}

The decision tree model is briefly introduced. The goal of a decision tree regression model is to predict $y$, the green time of a phase or the length of a barrier, based on a $d$-dimensional feature vector $\boldsymbol{X}=[x^1,x^2,...,x^d]^\mathsf{T}$, which are extracted from trajectories (see Section \ref{sec_critical} for more details on features). The training data consists of $n$ observations with their corresponding labels $\{y_1,y_2,...,y_n\}$ and features $\{\boldsymbol{X_1},\boldsymbol{X_2},...,\boldsymbol{X_n}\}$. During the learning process, the decision tree algorithm partitions the entire feature space into different sub-regions based on the training dataset. Figure \ref{tree} shows a simple example of a trained decision tree with two features. In this example, the first step is to divide the entire space into two sub-regions according to whether queue length is greater than 9. Similarly, in the second step, the left region is further divided into two sub-regions according to whether vehicle delay is greater than 12. Every step is called branching. For each sub-region, mean squared error (MSE), presented in Equation \ref{eq:1}, is calculated to evaluate the prediction performance of that branched sub-region based on the training data:
\begin{equation}\label{eq:1}
e_D = \sum_{j\in D} \frac{1}{|D|}(y_j-y_D)^2
\end{equation}
where $D$ is the set of all observations in this sub-region; $e_D$ is the MSE of set $D$; $|D|$ is the cardinality of set $D$; and $y_D$ is the mean value of the labels in set $D$.

Each time one performs branching, a node is split into two nodes. Each node represents a sub-region that contains data satisfying associated conditions. For each branching, the parent node is denoted as $D_p$ and the two child nodes are $D_{C1}$ and $D_{C2}$. Then the reduction in MSE due to a branching from a parent node into two child nodes, $\Delta I$, is defined as follows in Equation \ref{eq:2}: 
\begin{equation}\label{eq:2}
\Delta I = e_{D_p} - e_{D_{C1}} - e_{D_{C2}}
\end{equation}

All the observations in the parent data set are used as splitting candidates for the next level of branching. One can enumerate all the possible features and possible splits and calculate the corresponding $\Delta I$. The feature and the split with the maximum $\Delta I$ are chosen to branch the parent node. The branching process is repeated until the maximum number of iteration is reached.

For predictions of green time, one can start from the top of the tree, which is the root node, and find the path down to the final node, called the leaf node, based on the criteria of each node. The mean value of the labels in the final data set is the predicted green time. 

The process for the prediction of phase sequence is similar to that of green time. Because it is a classification problem, the majority vote of the labels is taken as the prediction.

\subsection{Critical Traffic Features}\label{sec_critical}
Different CV-TSC systems use different objectives and performance indexes to optimize the signal timing plan. Since the objectives are typically functions of one or more traffic features, the signal timing plan should be closely related to these associated traffic features, e.g., queue length (a signal controller allocates green time based on the queue length of each phase) and headway (a signal controller terminates a green phase when there is a large headway). A list of common traffic features applied in existing studies include queue length (QL) \citep{priemer2009decentralized}, number of approaching vehicles (NAV) \citep{goodall2013traffic}, headway (HW) \citep{koonce2008traffic}, estimated time of arrival (ETA) \citep{he2012pamscod,he2014multi}, vehicle delay (VD) \citep{wu2017delay}, and flow rate (FR) \citep{zheng2018traffic}.

For a particular CV-TSC system, not all traffic features are utilized to optimize the signal timing. We define the traffic features that determine the signal timing plan as critical features. When falsified data alter the values of these critical features, signal control decisions are changed accordingly. As a result, the attacker needs to identify the critical features that have a significant impact on the signal timing plan before launching attacks. Identifying critical features from the list of features is a feature selection problem. In this study, a sequential forward selection algorithm (SFS) is applied \citep{bow2002pattern}. Starting from an empty feature set, SFS greedily searches for the best features that can improve the prediction performance. \citet{john1994irrelevant}'s study suggests using SFS for identifying useful features and shows that SFS can improve the performance of decision tree models. The pseudo code of SFS is illustrated in Algorithm \ref{algorithm}. The output of SFS is a set that contains all the critical features. Note that in Section \ref{sec_tree}, only critical features are used as the input features of the decision tree models.

\begin{algorithm}\label{algorithm}
\SetAlgoLined
 Initialize two sets, $S$ and $R$. $S$ contains a complete list of features and $R$ is empty. Denote $e(Q)$ as the error when using feature subset $Q$. Initialize $e(R)$ to be a large number\;
 \While{$S$ is not empty}{
  find $s^*= \min_{s\in S} e(s\cup R)$\;
  \eIf{$e(s^*\cup R)<e(R)$}{
   Remove $s^*$ from $S$\;
   Add $s^*$ to $R$\;
   }{
   break while\;
  }
 }
 \KwResult{Set $R$ contains all the critical features}
 \caption{Sequential forward selection algorithm}
\end{algorithm}

\section{Case Study}\label{casestudy}
Through a case study in simulation, we realistically evaluate the impact of cyber attacks on a CV-TSC system under the black-box attack scenario. I-SIG system from the Multi-Modal Intelligent Traffic Signal System (MMITSS) project is selected as the targeted CV-TSCS system \citep{MMITSS}. Both simulation and field experiments have demonstrated the effectiveness of I-SIG in terms of delay reduction and mobility improvement \citep{feng2015real}.

\subsection{Control Logic of I-SIG}
The control logic of I-SIG system is briefly introduced in this subsection. For more details, please refer to \citet{feng2015real} and \citet{sen1997controlled}. At the beginning of each barrier, I-SIG takes a snapshot of the trajectories received from all the CVs within the RSU's communication range. Each trajectory is converted to ETA, which is calculated as the CV's distance to stop bar divided by its speed. Based on the ETAs, I-SIG solves a two-level optimization problem to find the optimal signal timing plan. At the lower level, I-SIG solves a utility minimization problem given the barrier length. The outputs of the lower level are the optimal green time and phase sequence. At the upper level, a dynamic programming (DP) problem is formulated with the objective to minimize total vehicle delay or total queue length. The decision variable at the upper level is the barrier length, which is considered as a stage in the DP formulation. Ideally, I-SIG should plan as many stages as needed so that all the vehicles can be properly served. For real-world implementations, however, I-SIG plans only two stages (i.e., one signal cycle) because of computational limitations in the RSU and real-time performance requirement. I-SIG then executes the timing plan of the first stage (the four phases in the current barrier) and arranges the phase sequence of the second stage (the four phases in the next barrier). When a new barrier starts, I-SIG repeats this optimization process.

\subsection{Simulation Setup}
A simulation environment is built using Matlab \citep{MATLAB:2018}. A typical 4-leg intersection with eight phases is modeled. Each approach has one left-turn lane and one through lane. For simplicity, right-turn lanes are not explicitly modeled. The car-following model from the NGSIM project is used to model vehicle motions \citep{yeo2008oversaturated}. The minimum green time and the maximum green time are set to be 5 seconds and 30 seconds for each phase, respectively. The transition time between phases (i.e., the yellow and red clearance time) is 4 seconds. The traffic demand for each movement is 400 vehicles per hour. The communication range is set to be 300 meters. The free flow speed is 15.65 meters per second (35 miles per hour). The resolution of the simulation is 10 Hz, which is consistent with the frequency of CV communication \citep{sae2009j2735}. Attacks are launched every time I-SIG optimizes the signal timing plan.

\subsection{Learning Control Logic}\label{learnlogic}
Because I-SIG only executes the optimized signal timing for one barrier each time, the surrogate model only needs to predict the green time of the four phases in the current barrier. Phase sequence can be obtained directly from SPaT messages. 

The surrogate model consists of two decision tree regression models. The output of the first decision tree model (labeled as Tree 1) is the barrier length (i.e., lead phase plus lag phase). The second decision tree model (labeled as Tree 2) outputs the green time of a lead phase. The green time of a lag phase is calculated by subtracting that of the corresponding lead phase from the barrier. The detailed definitions of the potential traffic features for the two decision tree models are shown in Table \ref{tbl:definition}. 

\begin{table}
\caption{Definitions of the traffic features associated with trees 1 and tree 2}
\label{tbl:definition}
\includegraphics[scale=0.75]{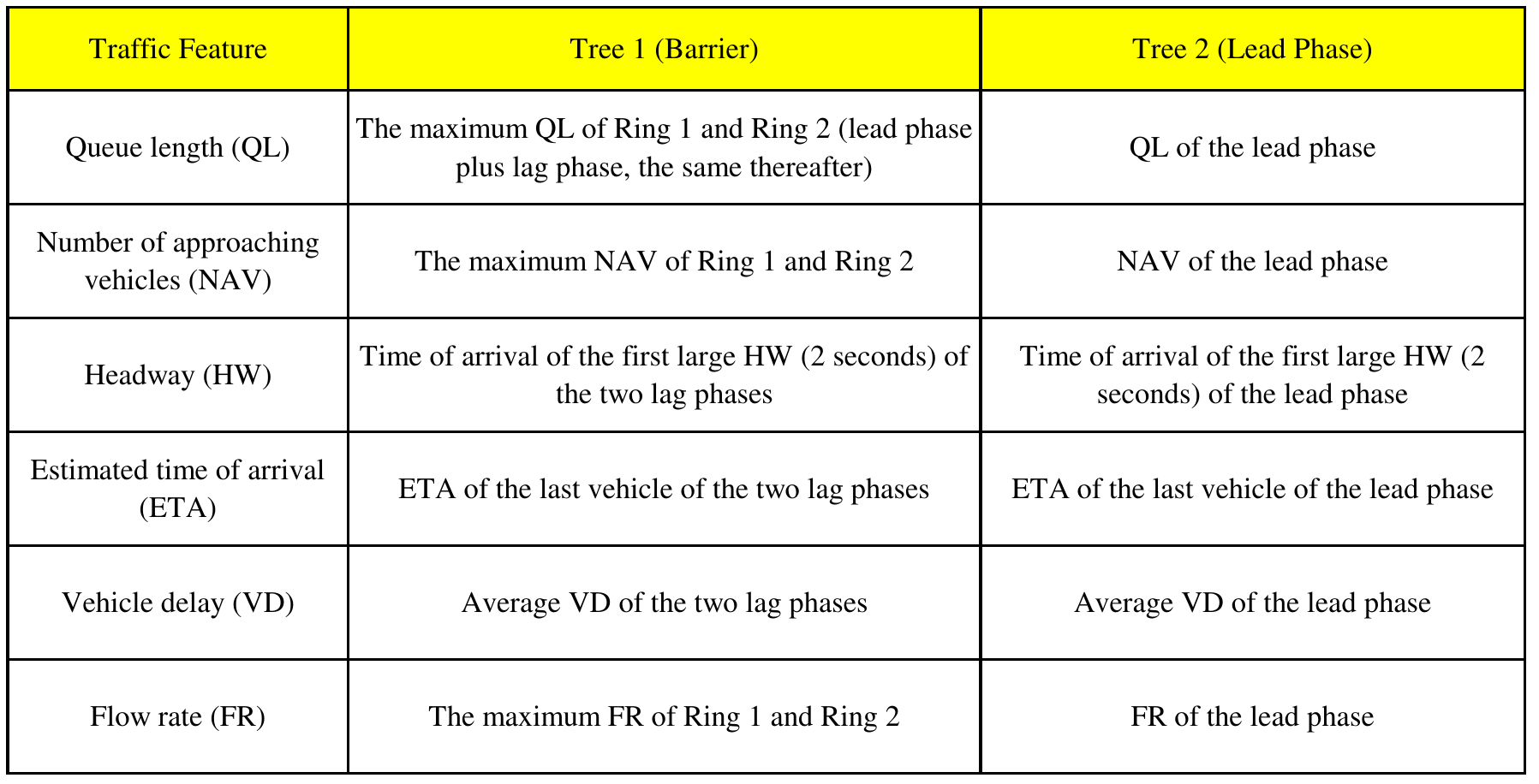}
\centering
\end{table}

A 30-hour simulation is run to generate a data set needed for both training and validation. Totally, 2206 optimizations are conducted. Mean absolute error (MAE), mean absolute percentage error (MAPE) and root mean square error (RMSE) are utilized to quantify errors for a given set of traffic features. Monte Carlo cross validation \citep{xu2001monte} is applied. 80\% of the data are randomly chosen for training, while the remaining 20\% are used for validation. The 80-20 process is repeated 10 times and the mean errors are recorded. The SFS is applied and the results are shown in Table \ref{tbl:att1}. In the first round, only one feature is used for fitting the decision tree models. The model with NAV has the least error for both trees. Therefore, NAV is added to the critical feature set. In the second round, two features are used for fitting the decision tree models, with one feature fixed to be NAV. The model with NAV and ETA has the least error. Thus, ETA is chosen as the second feature and added to the critical feature set. This process is repeated to find the third feature. However, the models with three features are all worse than the best model in the second round. Thus, SFS stops searching. NAV and ETA are identified as critical features. We note that the two identified critical features are consistent with the findings from a previous vulnerability analysis on I-SIG \citep{chen2018exposing}.

\begin{table}
\caption{Applying SFS for identifying critical traffic features}
\label{tbl:att1}
\includegraphics[scale=0.72]{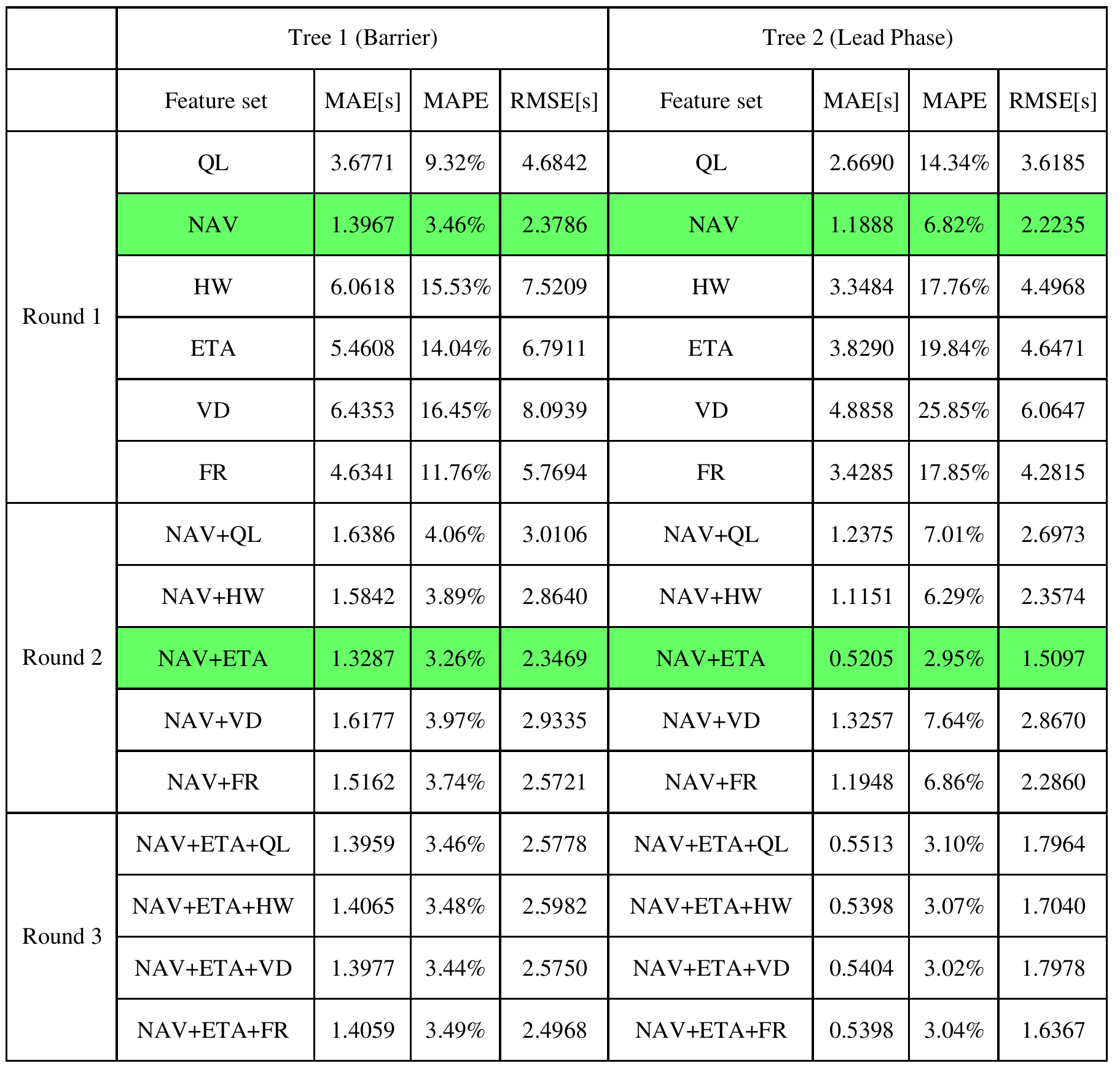}
\centering
\end{table}

\begin{figure}
\centering
\includegraphics[scale=0.7]{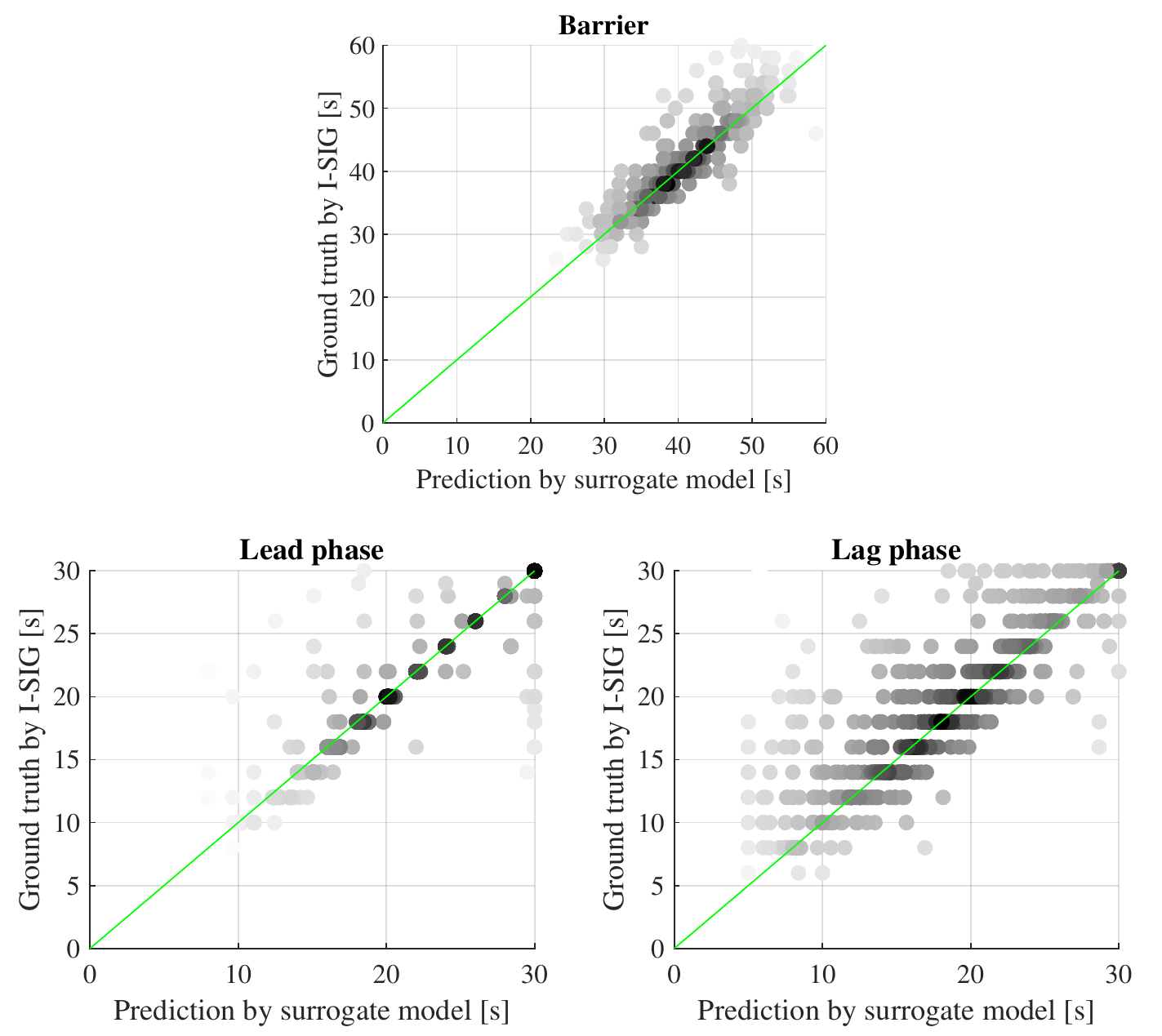}
\caption{Comparison between I-SIG and the trained surrogate model}
\label{flearning}
\end{figure}

Figure \ref{flearning} shows the effectiveness of the trained surrogate model. The prediction by the trained surrogate model and the ground truth generated by I-SIG are compared in the figure. The color depth represents the density of the data. The majority of the data lie on or near the 45-degree line, indicating that the surrogate model provides a good prediction accuracy. The prediction of the lag phase has a relatively greater error because it is estimated indirectly from the barrier and the lead phase.

\subsection{Evaluating the Impact of Cyber Attacks}\label{evaluating}
To evaluate the impact of cyber attacks, attacks are launched based on the assumed threat model. In Section \ref{learnlogic}, two critical features have been identified. Therefore, two attack cases are considered, with either ETA or NAV being the targeted critical feature. The prediction by the surrogate model (i.e., the predicted signal timing plan) includes the green time of the two lead phases and two lag phases in the current barrier, and is denoted as $\boldsymbol{f}(\boldsymbol{X}) = [g_{\text{d1}},g_{\text{d2}},g_{\text{g1}},g_{\text{g2}}]^\mathsf{T}$.

In the first case, the attacker is assumed to alter ETA by manipulating the location and speed data in the falsified BSMs. The attacker solves the following problem (\textbf{P2}) to maximize the dissimilarity between the pseudo-optimal timing plan and the timing plan under attack. 

\begin{align}
    \textbf{P2:} \quad & \nonumber \\
    \max \quad & \|\boldsymbol{f}(\boldsymbol{X_o})-\boldsymbol{f}(\boldsymbol{X_a})\|_2 \label{p2_1}\\
    \textrm{s.t.} \quad & \boldsymbol{X_o} = [t_{\text{d1}},t_{\text{d2}},t_{\text{g1}},t_{\text{g2}},n_{\text{d1}},n_{\text{d2}},n_{\text{g1}},n_{\text{g2}}]^\mathsf{T} \label{p2_2} \\
    & \boldsymbol{X_a} = [t_{\text{d1}}+\tau_{\text{d1}}\delta_{\text{d1}},t_{\text{d2}}+\tau_{\text{d2}}\delta_{\text{d2}},t_{\text{g1}}+\tau_{\text{g1}}\delta_{\text{g1}},t_{\text{g2}}+\tau_{\text{g2}}\delta_{\text{g2}}, \nonumber \\
    & n_{\text{d1}}+\delta_{\text{d1}},n_{\text{d2}}+\delta_{\text{d2}},n_{\text{g1}}+\delta_{\text{g1}},n_{\text{g2}}+\delta_{\text{g2}}]^\mathsf{T} \label{p2_3}\\
    & t_{\text{d1}}+\tau_{\text{d1}}\delta_{\text{d1}}, t_{\text{d2}}+\tau_{\text{d2}}\delta_{\text{d2}} \in T_{\text{lead}} \label{p2_6} \\ 
    & t_{\text{g1}}+\tau_{\text{g1}}\delta_{\text{g1}}, t_{\text{g2}}+\tau_{\text{g2}}\delta_{\text{g2}} \in T_{\text{lag}} \label{p2_7} \\ 
    & \delta_{\text{d1}} + \delta_{\text{d2}} + \delta_{\text{g1}} + \delta_{\text{g2}} = 1 \label{p2_8} \\
    & \tau_{\text{d1}},\tau_{\text{d2}},\tau_{\text{g1}},\tau_{\text{g2}} \geq 0 \label{p2_9} \\
    & \delta_{\text{d1}},\delta_{\text{d2}},\delta_{\text{g1}},\delta_{\text{g2}} \in \{0, 1\} \label{p2_10} \\
    \textrm{Decision variables} \quad & \tau_{\text{d1}},\tau_{\text{d2}},\tau_{\text{g1}},\tau_{\text{g2}},  \delta_{\text{d1}},\delta_{\text{d2}},\delta_{\text{g1}},\delta_{\text{g2}}
\end{align}

The observed critical features $\boldsymbol{X_o}$ include the ETAs of the two lead phases ($t_{\text{d1}},t_{\text{d2}}$) and two lags phases ($t_{\text{g1}},t_{\text{g2}}$) as well as the NAVs of the two lead phases ($n_{\text{d1}},n_{\text{d2}}$) and two lag phases ($n_{\text{g1}},n_{\text{g2}}$) (Equation \ref{p2_2}). Similarly, the altered critical features $\boldsymbol{X_a}$ also include corresponding ETAs and NAVs (Equation \ref{p2_3}). The binary variables $\delta_{\text{d1}},\delta_{\text{d2}},\delta_{\text{g1}},\delta_{\text{g2}}$ indicate the phase to which the falsified trajectory is injected (Equation \ref{p2_10}). We assume that the attacker can only inject one falsified trajectory per attack (Equation \ref{p2_8}). $T_{\text{lead}}$ and $T_{\text{lag}}$ are the sets of candidate ETAs for the lead phase and lag phase (Equation \ref{p2_6} and \ref{p2_7}). $T_{\text{lead}}$ and $T_{\text{lag}}$ can be obtained from historical data. Equation \ref{p2_9} indicates that the attacker intends to launch attacks by increasing the ETA. 

In the second case, the attacker is assumed to alter NAV by injecting falsified trajectories to different phases. Denote $\delta_{\text{d1}},\delta_{\text{d2}},\delta_{\text{g1}},\delta_{\text{g2}}$ to be the number of falsified trajectories injected to each phase. The attacker is assumed to have a budget limit $B=10$ (Equation \ref{p3_4}), i.e., the maximum number of falsified trajectories. Similar to the previous case, the attacker solves the following problem (\textbf{P3}) to maximizes the dissimilarity.

\begin{align}
    \textbf{P3:} \quad & \nonumber \\
    \max \quad & \|\boldsymbol{f}(\boldsymbol{X_o})-\boldsymbol{f}(\boldsymbol{X_a})\|_2 \label{p3_1}\\
    \textrm{s.t.} \quad & \boldsymbol{X_o} = [t_{\text{d1}},t_{\text{d2}},t_{\text{g1}},t_{\text{g2}},n_{\text{d1}},n_{\text{d2}},n_{\text{g1}},n_{\text{g2}}]^\mathsf{T} \label{p3_2} \\
    & \boldsymbol{X_a} = [t_{\text{d1}},t_{\text{d2}},t_{\text{g1}},t_{\text{g2}}, \nonumber \\
    & n_{\text{d1}}+\delta_{\text{d1}},n_{\text{d2}}+\delta_{\text{d2}},n_{\text{g1}}+\delta_{\text{g1}},n_{\text{g2}}+\delta_{\text{g2}}]^\mathsf{T} \label{p3_3}\\
    & \delta_{\text{d1}} + \delta_{\text{d2}} + \delta_{\text{g1}} + \delta_{\text{g2}} \leq B \label{p3_4} \\
    & \delta_{\text{d1}},\delta_{\text{d2}},\delta_{\text{g1}},\delta_{\text{g2}} \in \mathbb{Z}^+ \label{p3_5} \\
    \textrm{Decision variables} \quad & \delta_{\text{d1}},\delta_{\text{d2}},\delta_{\text{g1}},\delta_{\text{g2}}
\end{align}

\begin{figure}
\centering
\includegraphics[scale=.65]{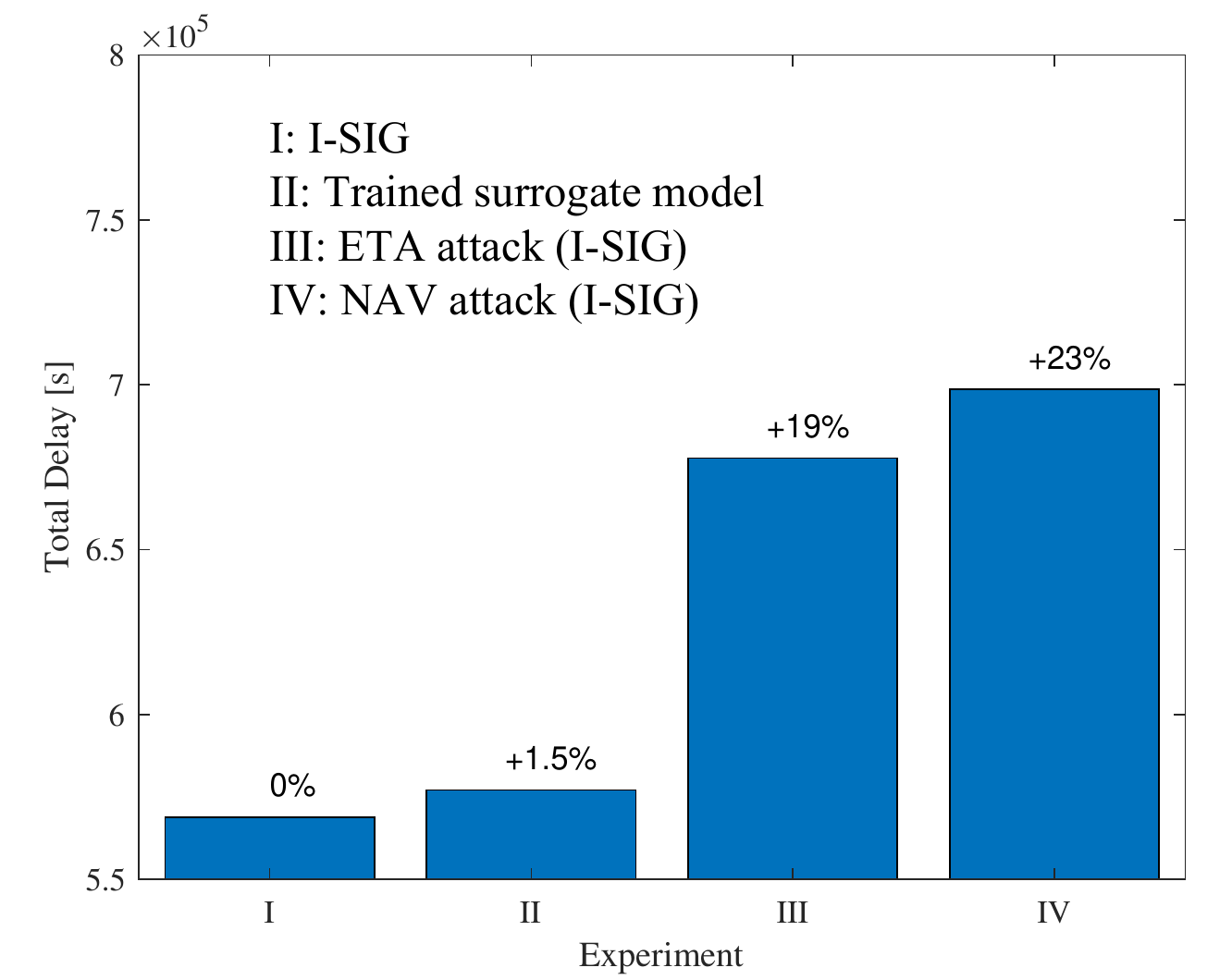}
\caption{Total delay for each experiment}
\label{exp}
\end{figure}

Four simulation experiments are conducted to assess the impact of cyber attacks. Each experiment lasts for 5 hours, with the exact same traffic demand and vehicle arrival patterns. The total delay for each experiment is shown in Figure \ref{exp}. 

In Experiment I, the original I-SIG system operates normally without being attacked. This experiment serves as the benchmark for all the other experiments. In Experiment II, the trained surrogate model is used to generate signal timing plans and control the traffic signal. However, no falsified CV data are injected to the system. As evidenced by a small increase of 1.5\%, the total delay is very close to the benchmark. This indicates that the trained surrogate model could effectively mimic the actual signal control logic. In Experiment III, the attacker is assumed to attack I-SIG based on the feature ETA. It is worth noting that only one falsified trajectory is injected into the system per attack. The total delay increases by 19\%. This implies that the attacker may cause significant damage to the CV-TSC system even without knowing the signal control logic. In Experiment IV, the attacker is assumed to attack I-SIG based on the feature NAV. Multiple falsified trajectories can be injected into the system per attack. The total delay increases by 23\%. When sufficient resources are given, the attacker may cause even more damage to the system. These experiment results show that the attacker may learn the signal control logic with a surrogate model. It is possible to attack a CV-TSC system without knowing the exact control logic. The falsified data attacks can result in severe consequences with a limited budget.

\subsection{Defense Strategy Discussion}
Although the focus of this study is on the impact evaluation of cyber attacks, defense strategies are briefly discussed. Because the attacker needs to learn the control logic before attack, one natural defense strategy is to interfere with the learning process. For example, we may add noise to the optimized signal timing plans. This may mislead the attacker's surrogate model. However, when the environment is benign (i.e., no attacks), adding noise makes the system operate under sub-optimal conditions. Thus, a trade-off needs to be made between security and efficiency.

Another defense strategy is to proactively identify falsified trajectories before they are utilized by the optimization model. The falsified trajectories are generated to alter the values of the critical features. For example, in our case study, to alter ETA to a large value, the attacker needs to generate a falsified trajectory which shows that the falsified vehicle approaches the intersection with a low speed. Such falsified trajectories are behaviorally different from normal trajectories. By monitoring abnormal behavior, we may be able to identify abnormal (falsified) trajectories. We will investigate this defense strategy in our future research.

\section{Conclusion}\label{conclusion}
In this study, we focus on evaluating the impact of cyber attacks on CV-TSC systems. Compared to the literature, this study considers a more realistic attack scenario in which the control logic of CV-TSC systems is unavailable to the attacker. We assume that the attacker may learn the signal control logic using a surrogate model and identify critical traffic features. With the learned model, the signal timing plan can be predicted by the attacker based on the identified critical traffic features. Falsified BSMs can be generated by the attacker to alter the values of the critical traffic features to launch real-time attacks. Consequently, the signal control decisions are influenced. The attacker is assumed to find the `optimal' value of the critical features by maximizing the dissimilarity between the pseudo-optimal timing plan and the resultant signal timing plan under attack. We evaluate the impact by comparing the system performance with and without attacks.

A comprehensive case study is conducted with I-SIG as the selected CV-TSC system. We find that the surrogate model can effectively mimic the actual control logic of I-SIG, which is sensitive to two critical traffic features: ETA (estimated time of arrival) and NAV (number of approaching vehicles). Two types of attacks are launched based on these two features. Simulation experiments show that the total delay increases by 19\% and 23\%. This indicates that even though the control logic is unknown, an attacker is still able to cause severe damage to the CV-TSC system. To protect the system, two defense strategies are briefly discussed. 

Based on the findings of this paper, future work will focus on the proactive defense strategy to safeguard CV-TSC systems from cyber attacks. The goal of the defense strategy is to detect anomaly in BSMs and filter out falsified trajectories before using them in any CV-based applications.

\section*{Funding}
This research is supported in part by the U.S. National Science Foundation through Grant SaTC \#1930041 and Mcity at the University of Michigan for financial support. The views presented in this paper are those of the authors alone.

\bibliographystyle{tfcad}
\bibliography{mybib}

\begin{thebibliography}{49}
\newcommand{\enquote}[1]{``#1''}
\providecommand{\natexlab}[1]{#1}
\providecommand{\url}[1]{\normalfont{#1}}
\providecommand{\urlprefix}{}

\bibitem[Beak, Head, and Feng(2017)]{beak2017adaptive}
Beak, Byungho, K~Larry Head, and Yiheng Feng. 2017. ``Adaptive coordination
  based on connected vehicle technology.'' \emph{Transportation Research
  Record} 2619 (1): 1--12.

\bibitem[Bow(2002)]{bow2002pattern}
Bow, Sing-Tze. 2002. \emph{Pattern recognition and image preprocessing}. Marcel
  Dekker New York.

\bibitem[Brecht et~al.(2018)]{brecht2018security}
Brecht, Benedikt, Dean Therriault, Andr{\'e} Weimerskirch, William Whyte,
  Virendra Kumar, Thorsten Hehn, and Roy Goudy. 2018. ``A Security Credential
  Management System for V2X Communications.'' \emph{IEEE Transactions on
  Intelligent Transportation Systems} 19 (12): 3850--3871.

\bibitem[Breiman(2017)]{breiman2017classification}
Breiman, Leo. 2017. \emph{Classification and regression trees}. Routledge.

\bibitem[Checkoway et~al.(2011)]{checkoway2011comprehensive}
Checkoway, Stephen, Damon McCoy, Brian Kantor, Danny Anderson, Hovav Shacham,
  Stefan Savage, Karl Koscher, et~al. 2011. ``Comprehensive experimental
  analyses of automotive attack surfaces.'' In \emph{USENIX Security
  Symposium}, Vol.~4. San Francisco.

\bibitem[Chen et~al.(2018)]{chen2018exposing}
Chen, Qi~Alfred, Yucheng Yin, Yiheng Feng, Z~Morley Mao, and Henry~X Liu. 2018.
  ``Exposing Congestion Attack on Emerging Connected Vehicle based Traffic
  Signal Control.'' In \emph{Network and Distributed Systems Security (NDSS)
  Symposium}, .

\bibitem[Ernst and Michaels(2017)]{ernst2017framework}
Ernst, Joseph~M, and Alan~J Michaels. 2017. ``Framework for evaluating the
  severity of cybervulnerability of a traffic cabinet.'' \emph{Transportation
  Research Record} 2619 (1): 55--63.

\bibitem[Feng et~al.(2015)]{feng2015real}
Feng, Yiheng, K~Larry Head, Shayan Khoshmagham, and Mehdi Zamanipour. 2015. ``A
  real-time adaptive signal control in a connected vehicle environment.''
  \emph{Transportation Research Part C: Emerging Technologies} 55: 460--473.

\bibitem[Feng et~al.(2018)]{feng2018vulnerability}
Feng, Yiheng, Shihong Huang, Qi~Alfred Chen, Henry~X Liu, and Z~Morley Mao.
  2018. ``Vulnerability of traffic control system under cyberattacks with
  falsified data.'' \emph{Transportation research record} 2672 (1): 1--11.

\bibitem[Feng, Yu, and Liu(2018)]{feng2018spatiotemporal}
Feng, Yiheng, Chunhui Yu, and Henry~X Liu. 2018. ``Spatiotemporal intersection
  control in a connected and automated vehicle environment.''
  \emph{Transportation Research Part C: Emerging Technologies} 89: 364--383.

\bibitem[Feng, Zheng, and Liu(2018)]{feng2018real}
Feng, Yiheng, Jianfeng Zheng, and Henry~X Liu. 2018. ``Real-Time Detector-Free
  Adaptive Signal Control with Low Penetration of Connected Vehicles.''
  \emph{Transportation Research Record} 2672 (18): 35--44.

\bibitem[Ghafouri et~al.(2016)]{ghafouri2016vulnerability}
Ghafouri, Amin, Waseem Abbas, Yevgeniy Vorobeychik, and Xenofon Koutsoukos.
  2016. ``Vulnerability of fixed-time control of signalized intersections to
  cyber-tampering.'' In \emph{2016 Resilience Week (RWS)}, 130--135. IEEE.

\bibitem[Ghena et~al.(2014)]{ghena2014green}
Ghena, Branden, William Beyer, Allen Hillaker, Jonathan Pevarnek, and J~Alex
  Halderman. 2014. ``Green lights forever: Analyzing the security of traffic
  infrastructure.'' In \emph{8th $\{$USENIX$\}$ Workshop on Offensive
  Technologies ($\{$WOOT$\}$ 14)}, .

\bibitem[Goodall, Smith, and Park(2013)]{goodall2013traffic}
Goodall, Noah~J, Brian~L Smith, and Byungkyu Park. 2013. ``Traffic signal
  control with connected vehicles.'' \emph{Transportation Research Record} 2381
  (1): 65--72.

\bibitem[Guler, Menendez, and Meier(2014)]{guler2014using}
Guler, S~Ilgin, Monica Menendez, and Linus Meier. 2014. ``Using connected
  vehicle technology to improve the efficiency of intersections.''
  \emph{Transportation Research Part C: Emerging Technologies} 46: 121--131.

\bibitem[He, Head, and Ding(2012)]{he2012pamscod}
He, Qing, K~Larry Head, and Jun Ding. 2012. ``PAMSCOD: Platoon-based arterial
  multi-modal signal control with online data.'' \emph{Transportation Research
  Part C: Emerging Technologies} 20 (1): 164--184.

\bibitem[He, Head, and Ding(2014)]{he2014multi}
He, Qing, K~Larry Head, and Jun Ding. 2014. ``Multi-modal traffic signal
  control with priority, signal actuation and coordination.''
  \emph{Transportation Research Part C: Emerging Technologies} 46: 65--82.

\bibitem[Jeske(2013)]{jeske2013floating}
Jeske, Tobias. 2013. ``Floating car data from smartphones: What google and waze
  know about you and how hackers can control traffic.'' \emph{Proc. of the
  BlackHat Europe} 1--12.

\bibitem[John, Kohavi, and Pfleger(1994)]{john1994irrelevant}
John, George~H, Ron Kohavi, and Karl Pfleger. 1994. ``Irrelevant features and
  the subset selection problem.'' In \emph{Machine Learning Proceedings 1994},
  121--129. Elsevier.

\bibitem[Kenney(2011)]{kenney2011dedicated}
Kenney, John~B. 2011. ``Dedicated short-range communications (DSRC) standards
  in the United States.'' \emph{Proceedings of the IEEE} 99 (7): 1162--1182.

\bibitem[Koonce and Rodegerdts(2008)]{koonce2008traffic}
Koonce, Peter, and Lee Rodegerdts. 2008. \emph{Traffic signal timing manual}.
  Technical {R}eport. United States Federal Highway Administration.

\bibitem[Koscher et~al.(2010)]{koscher2010experimental}
Koscher, Karl, Alexei Czeskis, Franziska Roesner, Shwetak Patel, Tadayoshi
  Kohno, Stephen Checkoway, Damon McCoy, et~al. 2010. ``Experimental security
  analysis of a modern automobile.'' In \emph{2010 IEEE Symposium on Security
  and Privacy}, 447--462. IEEE.

\bibitem[Laszka et~al.(2016)]{laszka2016vulnerability}
Laszka, Aron, Bradley Potteiger, Yevgeniy Vorobeychik, Saurabh Amin, and
  Xenofon Koutsoukos. 2016. ``Vulnerability of transportation networks to
  traffic-signal tampering.'' In \emph{2016 ACM/IEEE 7th International
  conference on Cyber-Physical Systems (ICCPS)}, 1--10. IEEE.

\bibitem[Lee, Park, and Yun(2013)]{lee2013cumulative}
Lee, Joyoung, Byungkyu Park, and Ilsoo Yun. 2013. ``Cumulative travel-time
  responsive real-time intersection control algorithm in the connected vehicle
  environment.'' \emph{Journal of Transportation Engineering} 139 (10):
  1020--1029.

\bibitem[Li and Ban(2018)]{li2018connected}
Li, Wan, and Xuegang Ban. 2018. ``Connected Vehicles Based Traffic Signal
  Timing Optimization.'' \emph{IEEE Transactions on Intelligent Transportation
  Systems} .

\bibitem[MATLAB(2018)]{MATLAB:2018}
MATLAB. 2018. \emph{9.7.0.1190202 (R2019b)}. Natick, Massachusetts: The
  MathWorks Inc.

\bibitem[Mazloom et~al.(2016)]{mazloom2016security}
Mazloom, Sahar, Mohammad Rezaeirad, Aaron Hunter, and Damon McCoy. 2016. ``A
  security analysis of an in-vehicle infotainment and app platform.'' In
  \emph{10th $\{$USENIX$\}$ Workshop on Offensive Technologies ($\{$WOOT$\}$
  16)}, .

\bibitem[Oza et~al.(2020)]{oza2020secure}
Oza, Pratham, Mahsa Foruhandeh, Ryan Gerdes, and Thidapat Chantem. 2020.
  ``Secure Traffic Lights: Replay Attack Detection for Model-based Smart
  Traffic Controllers.'' In \emph{Proceedings of the Second ACM Workshop on
  Automotive and Aerial Vehicle Security}, 5--10.

\bibitem[Pandit et~al.(2013)]{pandit2013adaptive}
Pandit, Kartik, Dipak Ghosal, H~Michael Zhang, and Chen-Nee Chuah. 2013.
  ``Adaptive traffic signal control with vehicular ad hoc networks.''
  \emph{IEEE Transactions on Vehicular Technology} 62 (4): 1459--1471.

\bibitem[Papernot et~al.(2017)]{papernot2017practical}
Papernot, Nicolas, Patrick McDaniel, Ian Goodfellow, Somesh Jha, Z~Berkay
  Celik, and Ananthram Swami. 2017. ``Practical black-box attacks against
  machine learning.'' In \emph{Proceedings of the 2017 ACM on Asia conference
  on computer and communications security}, 506--519.

\bibitem[Perrine et~al.(2019)]{perrine2019implications}
Perrine, Kenneth~A, Michael~W Levin, Cesar~N Yahia, Melissa Duell, and
  Stephen~D Boyles. 2019. ``Implications of traffic signal cybersecurity on
  potential deliberate traffic disruptions.'' \emph{Transportation research
  part A: policy and practice} 120: 58--70.

\bibitem[Priemer and Friedrich(2009)]{priemer2009decentralized}
Priemer, Christian, and Bernhard Friedrich. 2009. ``A decentralized adaptive
  traffic signal control using V2I communication data.'' In \emph{2009 12th
  International IEEE Conference on Intelligent Transportation Systems}, 1--6.
  IEEE.

\bibitem[Reilly et~al.(2016)]{reilly2016creating}
Reilly, Jack, S{\'e}bastien Martin, Mathias Payer, and Alexandre~M Bayen. 2016.
  ``Creating complex congestion patterns via multi-objective optimal freeway
  traffic control with application to cyber-security.'' \emph{Transportation
  Research Part B: Methodological} 91: 366--382.

\bibitem[SAE(2009)]{sae2009j2735}
SAE, Draft. 2009. ``J2735 dedicated short range communications (dsrc) message
  set dictionary.'' \emph{Society of Automotive Engineers, DSRC Committee} .

\bibitem[Sen and Head(1997)]{sen1997controlled}
Sen, Suvrajeet, and K~Larry Head. 1997. ``Controlled optimization of phases at
  an intersection.'' \emph{Transportation science} 31 (1): 5--17.

\bibitem[Sinai et~al.(2014)]{sinai2014exploiting}
Sinai, Meital~Ben, Nimrod Partush, Shir Yadid, and Eran Yahav. 2014.
  ``Exploiting social navigation.'' \emph{arXiv preprint arXiv:1410.0151} .

\bibitem[USDOT(2019{\natexlab{a}})]{CVpilot}
USDOT. 2019{\natexlab{a}}. ``Connected vehicle pilot deployment program.''
  \url{https://www.its.dot.gov/pilots/}. Online; Accessed: 2019-05-21.

\bibitem[USDOT(2019{\natexlab{b}})]{MMITSS}
USDOT. 2019{\natexlab{b}}. ``Multi-Modal Intelligent Traffic Safety System.''
  \url{https://www.its.dot.gov/research_archives/dma/bundle/mmitss_plan.htm}.
  Online; Accessed: 2019-05-21.

\bibitem[Wang, Mao, and Gong(2017)]{wang2017overview}
Wang, Xuyu, Shiwen Mao, and Michelle~X Gong. 2017. ``An overview of 3GPP
  cellular vehicle-to-everything standards.'' \emph{GetMobile: Mobile Computing
  and Communications} 21 (3): 19--25.

\bibitem[Whyte et~al.(2013)]{whyte2013security}
Whyte, William, Andr{\'e} Weimerskirch, Virendra Kumar, and Thorsten Hehn.
  2013. ``A security credential management system for V2V communications.'' In
  \emph{2013 IEEE Vehicular Networking Conference}, 1--8. IEEE.

\bibitem[Wu et~al.(2017)]{wu2017delay}
Wu, Jian, Dipak Ghosal, Michael Zhang, and Chen-Nee Chuah. 2017. ``Delay-based
  traffic signal control for throughput optimality and fairness at an isolated
  intersection.'' \emph{IEEE Transactions on Vehicular Technology} 67 (2):
  896--909.

\bibitem[Xu and Liang(2001)]{xu2001monte}
Xu, Qing-Song, and Yi-Zeng Liang. 2001. ``Monte Carlo cross validation.''
  \emph{Chemometrics and Intelligent Laboratory Systems} 56 (1): 1--11.

\bibitem[Yang et~al.(2019)]{yang2019eco}
Yang, Zhen, Yiheng Feng, Xun Gong, Ding Zhao, and Jing Sun. 2019.
  ``Eco-trajectory Planning with Consideration of Queue Along Congested
  Corridor for Hybrid Electric Vehicles.'' \emph{Transportation Research
  Record} 2673 (9): 277--286.

\bibitem[Yen et~al.(2018)]{yen2018falsified}
Yen, Chia-Cheng, Dipak Ghosal, Michael Zhang, Chen-Nee Chuah, and Hao Chen.
  2018. ``Falsified Data Attack on Backpressure-based Traffic Signal Control
  Algorithms.'' In \emph{2018 IEEE Vehicular Networking Conference (VNC)},
  1--8. IEEE.

\bibitem[Yeo et~al.(2008)]{yeo2008oversaturated}
Yeo, Hwasoo, Alexander Skabardonis, John Halkias, James Colyar, and Vassili
  Alexiadis. 2008. ``Oversaturated freeway flow algorithm for use in next
  generation simulation.'' \emph{Transportation Research Record} 2088 (1):
  68--79.

\bibitem[Yu et~al.(2018)]{yu2018integrated}
Yu, Chunhui, Yiheng Feng, Henry~X Liu, Wanjing Ma, and Xiaoguang Yang. 2018.
  ``Integrated optimization of traffic signals and vehicle trajectories at
  isolated urban intersections.'' \emph{Transportation research part B:
  methodological} 112: 89--112.

\bibitem[Yu et~al.(2019)]{yu2019corridor}
Yu, Chunhui, Yiheng Feng, Henry~X Liu, Wanjing Ma, and Xiaoguang Yang. 2019.
  ``Corridor level cooperative trajectory optimization with connected and
  automated vehicles.'' \emph{Transportation Research Part C: Emerging
  Technologies} 105: 405--421.

\bibitem[Zhao and Chen(2017)]{zhao2017location}
Zhao, Bo, and Qinying Chen. 2017. ``Location spoofing in a location-based game:
  A case study of Pok{\'e}mon Go.'' In \emph{International Cartographic
  Conference}, 21--32. Springer.

\bibitem[Zheng et~al.(2018)]{zheng2018traffic}
Zheng, Jianfeng, Weili Sun, Shihong Huang, Shengyin Shen, Chunhui Yu, Jinqing
  Zhu, Bingbing Liu, and Henry~X Liu. 2018. \emph{Traffic Signal Optimization
  Using Crowdsourced Vehicle Trajectory Data}. Technical {R}eport.
  Transportation Research Board.

\end{thebibliography}

\end{document}